\documentclass[aip,apl,reprint]{revtex4-1}
\usepackage{graphicx}
\usepackage{hyperref}
\usepackage{bm}
\usepackage{amsmath} 
\usepackage{systeme}
\usepackage{comment}
\usepackage{xcolor}
\definecolor{newtext}{RGB}{0, 0, 0}
\definecolor{newtext2}{RGB}{0, 0, 0}

\begin{document}

\title{Spectrum evolution and chirping of laser-induced spin wave packets in thin iron films}

\author{Ia. A. Filatov}
\email[]{yaroslav.filatov@mail.ioffe.ru}
\homepage[]{http://www.ioffe.ru/ferrolab/}

\author{P. I. Gerevenkov}
\affiliation{Ioffe Institute, 26 Politekhnicheskaya st., 194021, St. Petersburg, Russian Federation}
\author{M. Wang}
\author{A. W. Rushforth}
\affiliation{School of Physics and Astronomy, The University of Nottingham, NG7 2RD Nottingham, UK}
\author{A. M. Kalashnikova}
\author{N. E. Khokhlov}
\affiliation{Ioffe Institute, 26 Politekhnicheskaya st., 194021, St. Petersburg, Russian Federation}

\date{\today}

\begin{abstract}
We present an experimental study of ultrafast optical excitation of magnetostatic \textcolor{newtext}{surface} spin wave (MSSW) packets and their spectral properties in thin films of pure iron. 
As the packets leave the excitation area and propagate in space, their spectra evolve non-trivially.
Particularly, low or high frequency components are suppressed at the border of the excitation area depending on the orientation of the external magnetic field with respect to the magnetocrystolline anisotropy axes of the film.
The effect is ascribed to the ultrafast local heating of the film.
Further, the time resolution of the implemented all-optical technique allows us to extract the chirp of the MSSW packet in the time domain via wavelet analysis.
The chirp is a result of the group velocity dispersion of the MSSW and, thus, is controlled by the film's magnetic parameters, magnetization and anisotropy, and external field orientation.
The demonstrated tunable modulation of MSSW wave packets with femtosecond laser pulses may find application in future magnonic-photonic hybrid devices for wave-based data processing. 
\end{abstract}


\maketitle

Developments in neuromorphic computing and pattern recognition largely rely on signal processing, going beyond the schemes offered by conventional electronics. 
Spin waves (SWs), or coherent magnons, in magnetically ordered structures are considered as a potential alternative to electrons in semiconductor devices for the transmission and processing of information in a special area of spintronics named magnonics  \cite{Pirro_advances_in_magnonics_NatRevMaterials2021, bookChapter_Spintronics_2021, Nikitov:UFN2015, Mahmoud_JAP_2020_Intro_to_SW_computing, Barman_Magnonic_RoadMap_2021}.
The most prominent feature of SWs is the possibility to transfer information about the magnetic moment without electric currents, thereby minimizing ohmic losses.
\textcolor{newtext}{It stimulates} a growing number of works \textcolor{newtext}{addressing} to SWs as an effective tool for wave-based logic and \textcolor{newtext}{even non-Boolean} calculations \cite{Khitun_Magnonic_logic_JPhysD2010, Csaba_perspectives_SWcomp_PLA_2017, Khivintsev_spin_logic_PMM2019, Mahmoud_JAP_2020_Intro_to_SW_computing, Barman_Magnonic_RoadMap_2021}.
Moreover, SW transport is \textcolor{newtext}{already demonstrated at sub-100-nm scales in magnonic conduits\cite{Wang_SpinPinning_PRL2019, Heinz_Chumak_SWpackets-in-nm-YIG-waveguide_NanoLetters2020} as well as the controllable interaction of SWs with nanometer sized natural magnetic textures\cite{Wagner_DWasSWwaveguide_NatNanotech2016, Huajun_ElectriControlofSW_AdvMaterials2021, Li_SWin2DVortexNetwork_NanoLett2021, LanXiao_SWpackets_and_skyrmions}.}
However, such a small scale for real applications demands appropriate materials with easy-to-manufacture procedures and satisfactory magnonic properties, such as high frequencies of the SWs, their tunability and easy integration into semiconductor electronic circuits.
These features are native for transition-metal based micro- and nanometer sized waveguides for SWs, making them promising candidates for nano-magnonic applications\cite{bookChapter_Spintronics_2021, Mahmoud_JAP_2020_Intro_to_SW_computing}. 
Furthermore, the intrinsic strong anisotropy fields in metal\textcolor{newtext}{lic} nano-waveguides play the role of an additional knob to design the properties of SWs, such as the SW dispersion and the corresponding phase evolution \cite{Sekiguchi_NPG_Asia_mat_2017, khokhlov-PRAppl2019}.
The latter is crucial to make SWs effective tools for wave-based logic and interferometric calculations\cite{Khitun_Magnonic_logic_JPhysD2010, Csaba_perspectives_SWcomp_PLA_2017, Khivintsev_spin_logic_PMM2019, Mahmoud_JAP_2020_Intro_to_SW_computing, Barman_Magnonic_RoadMap_2021}, particularly, in speeding up computationally demanding tasks such as image and speech recognition \textcolor{newtext}{based on magnonic neural networks} \cite{Kozhevnikov_pattern_recognition_APL2015, Albisetti_Nanomagnonics_AdvMat2020, Papp_NeuroMorficSW_NatComm2021}.
The described practical purposes of magnonics require appropriate research tools to inspect the wave nature of SWs.
Recently, ultrafast laser pulses were introduced as such a tool because they can excite SWs with exceptional tunability \cite{Satoh_NatPhotonics:2012, Chernov_SpinWavePhase_PhotResearch2018, Yoshimine_Unidirectional_control_EPL2017, Hioki_bi-reflection_CommPhys2020, khokhlov-PRAppl2019,Au_Direct_Excitation_SW_PhysRevLett2013, KHOKHLOV_JMMM2021}.
However, laser excitation is intrinsically broad-band and inhomogenious, which has a significant effect on the properties of the generated SW packets, limiting their spatial and temporal profile and leading to large spectral width.
Therefore, understanding the propagation of such wave packets is essential for developing applications of optically generated SWs for future magnonic-photonic hybrid devices.

In this letter, we report on the experimental observation of controllable spectrum evolution and frequency modulation of laser-induced magnetostatic surface waves (MSSWs) in thin films of pure iron.
We show, that the spectral features of MSSW wave packets can be tuned by choosing the orientation of an external magnetic field with respect to the axes of the magnetocrystalline anisotropy.
As the optically excited MSSW packets are broad-band and propagate in a dispersive medium, they obtain a chirp --- a time delay between distinct frequency components.
We clearly detect the chirping at the $\mu$m spatial scale.
The findings are crucial for the design of hybrid opto-magnonic wave-based data processing circuits.

 \begin{figure}[b]
 \includegraphics[width=1\linewidth]{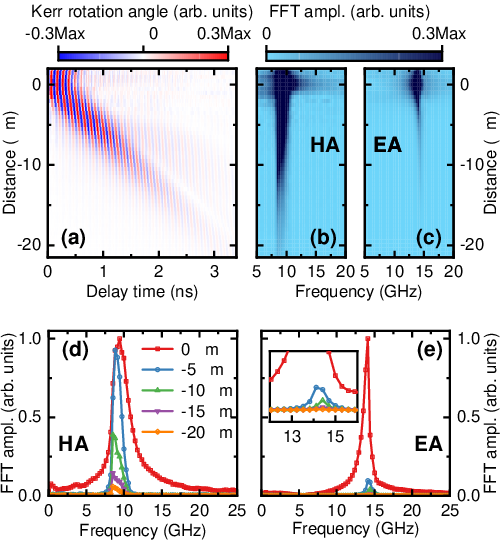}
 \caption{\label{fig:10nm_experiment} (a) Polar Kerr signal induced by propagating MSSW packet in the time-space domain for Fe film with $d = 10$ nm in the HA-configuration.
 (b,c) Space-frequency maps of the packet for the HA- (b) and EA- (c) configurations.
 (d, e) FFT spectra of the packet at different pump-probe distances for the HA- (d) and EA- (e) configurations.
 Inset in (e) shows enlarged area of the same spectra.}
 \end{figure}

Our investigations are carried out using epitaxial films of iron with thicknesses $d$ = 10 and 20 nm.
The films are deposited on 350-$\mu$m thick GaAs (001) substrates by magnetron sputtering and are capped with a 2.5-nm-thick chromium layer for protective purposes. 
The films possess pronounced cubic and interface-induced uniaxial in-plane magnetic anisotropy similar to other epitaxial thin films of iron and its alloys on GaAs substrates \cite{Gester1996, Wastlbauer_Fe_AdvPhys2005, Parkes_magnetostrictive_SciRep2013}.

Excitation and detection of MSSWs are performed using \textcolor{newtext}{the setup of} space-time-resolved magneto-optical Kerr effect  \cite{Filatov_spectrum_evolution_JPCS2020}.
Pump and probe laser pulses with central wavelengths of 750 nm and 1050 nm, respectively, a duration of 120 fs, 70 MHz repetition rate, are focused normally onto opposite sides of the iron film into 3.5-$\mu$m-diameter spots  using microobjectives.
The fluence of the pump pulses is 4.5 mJ/cm$^2$, and the probe pulses have a fluence about 20 times lower.
The temporal resolution is achieved with an opto-mechanical delay line placed in the pump path.
The spatial scan is realized by placing the pump microobjective on a piezo stage.
\textcolor{newtext}{An external static magnetic field is applied in the film’s plane.
Its magnitude $\mu_{0}{\bf H} = 100$ mT which is twice as large as the in-plane anisotropy field, provides alignment of the film's magnetization along {\bf H}.
Thus, varying the distance between the pump and probe spots perpendicularly to {\bf H} enables the detection of the surface mode of magnetostatic waves \cite{DAMON1961308}.}
The excitation mechanism of MSSWs is an ultrafast change of the magnetocrystalline anisotropy caused by ultrafast heating induced by the pump pulse, similar to the recent experiments of MSSW excitation in galfenol (Fe$_{0.81}$Ga$_{0.19}$) films \cite{khokhlov-PRAppl2019, Filatov_spectrum_evolution_JPCS2020}.
We investigate propagating MSSWs for two directions of {\bf H}: (i) {\bf H} is aligned along the sample's hard anisotropy axis and (ii) {\bf H} is at 15$^{\circ}$ with respect to the easy anisotropy axis, referred further as the HA- and EA-configuration, respectively.

Figure \ref{fig:10nm_experiment}(a) shows the spatial-temporal map of the polar Kerr rotation angle for the probe measured in the sample with $d = 10$ nm in the HA-configuration.
The color bar spans up to 0.3 of the maximum amplitude of the measured Kerr signal to show the propagation of the MSSW packet beyond the distance at which its amplitude decreases by a factor of \textit{e}.
The shape and slope of the phase pattern serves as evidence that the detected Kerr signal originates from the propagating MSSW wave packet.
Thus, the propagation of laser-induced MSSW wave packets over a distance of more than 10 $\mu$m is clearly observed. 
Such propagation distance is comparable and even larger than that of typical values in other metallic materials used in real magnonic circuits \cite{Chumak_magnonic_crystals_JPhysD2017}.

Next, we analyze the spectral composition of the propagating MSSW wave packet by performing fast Fourier transform (FFT) analysis of the signal in the time domain for each pump-probe distance.
The result is shown in Fig.\,\ref{fig:10nm_experiment}(b,c) for the 10-nm sample in the HA- and EA-configurations, respectively.
Compression of the spectral width and a shift of the frequency with maximal FFT amplitude are observed during the propagation of the MSSW \textcolor{newtext}{packet}.
More explicit confirmation of such spectral changes is possible by analyzing the individual spectra at different pump-probe distances (Fig.\,\ref{fig:10nm_experiment}(d,e)).
The central frequency of the propagating MSSW shifts to the lower or higher part of the spectrum observed in the excitation area for the HA- and EA-configurations, respectively, in agreement with previous results for galfenol films on GaAs substrates \cite{Filatov_spectrum_evolution_JPCS2020, khokhlov-PRAppl2019}.
The observed spectral evolution is associated with different MSSW dispersion relations inside and outside the pump-heated  area.
Indeed, effective anisotropy fields are partially suppressed by an abrupt decrease of the anisotropy parameters and magnetization inside the excitation area \cite{Gerevenkov_effect_anisotropy_PRMat2021}.
This leads to a shift of the MSSW dispersion inside the pump area to the higher or lower frequency range depending on the orientation of {\bf H} because of the different manner in which the anisotropy field enters the total effective field: i.e. the effective anisotropy field is subtracted from (added to) {\bf H} in the HA-(EA-)configuration, leading to a decrease (increase) of the total effective magnetic field in the sample, which defines the frequency of the MSSW \cite{Gurevich_book_1996, Kalinikos_SW_dispersion_JPhysCondMat1990}.
Since the pump has a limited diameter, it excites a limited band of wavenumbers inside the spot.
This sets an upper limit on the spectral range of the excited MSSW frequencies, with the lower limit being the ferromagnetic resonance frequency.
\textcolor{newtext}{As the result}, MSSWs propagating outside the pump have the same limitation on the wavenumber, but the corresponding frequency range is shifted \textcolor{newtext}{to the bottom (top) part of the initially excited spectrum.}

\textcolor{newtext}{It is worth mentioning that the spectral asymmetry of the precession inside the pump spot is attributed to another origin.
Particularly, the precession occurs in the abruptly heated trap for MSSWs  \cite{Kolokoltsev_HotResonator_APL2012, Busse_SWtraps_SciRep2015} with the temperature recovery on the order of several hundred picoseconds \cite{Gerevenkov_effect_anisotropy_PRMat2021, Busse_SWtraps_SciRep2015}.
The corresponding evolution of the magnetic parameters results in a gradual recovery of the pump-induced up-(down-)shift of the frequency back to its equilibrium value, giving rise to longer lifetimes and higher FFT amplitudes of the bottom (top) part of the spectrum in the HA-(EA-)configuration.}

Using FFT spectra, we determine values of the MSSW propagation length $l_p$ for both magnetic field configurations by fitting the dependence of the \textcolor{newtext}{FFT's integral}, $S$ on the pump-probe distance, $L$ with the exponential function \textcolor{newtext}{$S(L)= S(0) \mathrm{exp}(-L/l_p)$}.
We determine $l_p$ in the \textcolor{newtext}{region} of $L \geq 3$ $\mu$m, i.e. starting from the edge of the pump \textcolor{newtext}{spot}, which represents the value of $l_p$ for the MSSW propagating in the non-heated iron film.
The obtained values of $l_p$ for all samples are larger than 6\,$\mu$m in the HA-configuration and less than 5\,$\mu$m in the EA-configuration (Table \ref{table:Params_fFMR_F}).
The observation that $l_p$ is larger in the HA-configuration is in good agreement with previous works on MSSWs in cubic anisotropic metal films \cite{Sekiguchi_NPG_Asia_mat_2017, khokhlov-PRAppl2019}.
The effect is connected with the reduction of the total in-plane effective field in the HA-configuration \textcolor{newtext}{provided by the modification of the dispersion relation determined by the mutual orientation of the magnetic crystalline anisotropy axes and the direction of the SW's wave vector.
It results in an enhancement of SW's parameters such as amplitude, propagation length and group velocity in the HA-configuration as compared to the EA-configuration, which is discussed in detail elsewhere \cite{Sekiguchi_NPG_Asia_mat_2017}.}

\begin{table*}
\caption{\label{table:Params_fFMR_F}%
Parameters of MSSW obtained in experiments}
\begin{ruledtabular}
\begin{tabular}{c|ccc|ccc}
 d & \multicolumn{3}{c|}{HA-configuration}   &  \multicolumn{3}{c}{EA-configuration} \\
 (nm) & $l_p$ ($\mu$m) & $f_{FMR}$ (GHz) & $F$ (GHz) & $l_p$ ($\mu$m) & $f_{FMR}$ (GHz) & $F$ (GHz) \\
\colrule
 10 & $6.1 \pm 0.1$ & $8.3 \pm 0.1$ & $152 \pm 4$ & $2.6 \pm 0.1$ & $13.9 \pm 0.2$ & $220 \pm 10$  \\ 
 20 & $8.9 \pm 0.3$ & $8.2 \pm 0.1$ & $148 \pm 5$ & $4.7 \pm 0.1 $ & $14.9\pm 0.3$ & $150 \pm 15$  \\ 
\end{tabular}
\end{ruledtabular}
\end{table*}

\begin{figure}
\includegraphics[width=1 \linewidth]{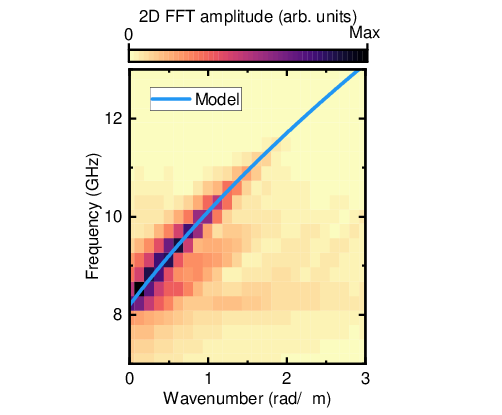}
 \caption{\label{Fig:Disp} Dispersion of MSSW wave packet in 20-nm iron film in \textcolor{newtext}{the} HA-configuration, reconstructed from experimental \textcolor{newtext}{spatial-temporal} maps (pseudo-color) and calculated analytically with Eq.\,(\ref{Eq:MSW_quad_disp}) (line).
 }
\end{figure}

\begin{figure}
 \includegraphics[width=1 \linewidth]{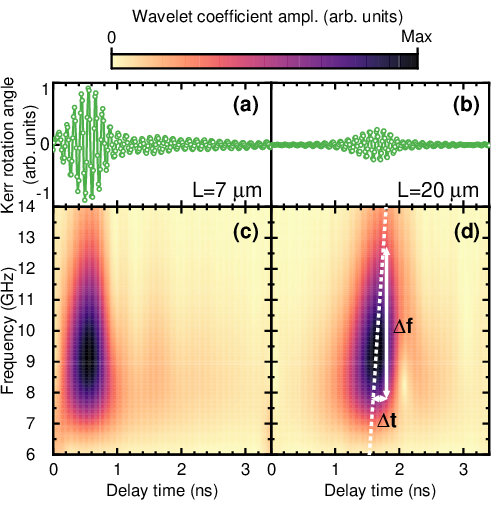}
 \caption{\label{Fig:Wavelet} (a,b) Wave packet signals at distances of 7 and 20 $\mu$m from the pump spot in the time domain for the 20-nm film in the HA-configuration.
 (c,d) Corresponding wavelet spectrograms in the time-frequency domain.
 Dashed line on (d) shows \textcolor{newtext}{the} variation of the frequency peak with time.
 Double arrows show the variation of the frequency $\Delta f$ and time shift $\Delta t$ schematically.}
 \end{figure}

\textcolor{newtext}{For further investigation of the wave packets we reconstruct the MSSW dispersion $f(k)$ from spatial-temporal maps via two-dimensional FFT (Fig.\,\ref{Fig:Disp}), where $k$ is the MSSW wavenumber.
The reconstructed dispersion has a form different from the linear one with $\frac{\partial^2 f}{\partial k^2}\neq0$.
It means the group velocity of the MSSW wavepacket has a nonzero dispersion, which should impose frequency modulation, or chirp, across the pulse during its propagation \cite{Agrawal_book_2019,Akhmanov_WavePackets_UFN1986}. 
Moreover, the increase of asymmetry in the spectra with $L$ (Fig.\,\ref{fig:10nm_experiment}(d,e)) could be the second trace of the chirp.
To reveal the chirping,} we implement continuous wavelet transform with Morlet mother wavelet similar to opto-acoustic pump-probe experiments \cite{Dehoux_non-invasive_APL_2012, Danworaphong_3D_bioImaging_APL2015}, \textcolor{newtext}{which allows the extraction of} information about the frequency evolution in the time domain.
Indeed, the resulting analysis reveals the shift of the frequency components $\Delta f$ with time $t$, increasing with $L$ (Fig.\,\ref{Fig:Wavelet}(c,d)).

To characterise the shift, we approximate the registered frequency modulation of the MSSW packets with a linear function $f(t) = f_c+\nu\textcolor{newtext}{(L)} t$, where $f_c$ is a wave packet central frequency, $\nu\textcolor{newtext}{(L)} = \frac{\partial f}{\partial t}$ is the chirp parameter.
\textcolor{newtext}{The theoretical dependency $\nu(L)$ follows from expanding the dispersion relation $k(f)$ into a Taylor series up to the derivative of the second order \cite{Agrawal_book_2019, Akhmanov_WavePackets_UFN1986, Zvezdin_NonLinearSW_JETP1983}.
The procedure is described in detail in Suppl. Materials.
Particularly,} to compare the \textcolor{newtext}{theoretical $\nu(L)$  with our experimental results,} we used the analytic dispersion of MSSWs in the thin film limit ($kd \ll 1$) and for small wavenumbers (the exchange contribution to the spin wave energy is negligible) in a simplified form \cite{Arias_FMR_PRB1999, Kalinikos_SW_dispersion_JPhysCondMat1990, Zakeri_magnonic_crustals_JPCondMat2020}:
\begin{equation}\label{Eq:MSW_quad_disp}
    f(k) = \left( f_{FMR}^2 + F^2 \frac{kd}{2}\right )^{1/2},
\end{equation}
where $f_{\text{FMR}}$ is the ferromagnetic resonance frequency, and $F$ is a constant.
\textcolor{newtext}{Theoretical} values of $f_{\text{FMR}}$ and $F$ are determined by the gyromagnetic ratio, saturation magnetization, and anisotropy parameters of the film, and the mutual orientation of {\bf H} and anisotropy axes.
Here we obtain $f_{\text{FMR}}$ and $F$ by fitting the experimental MSSW dispersion reconstructed from spatial-temporal maps (Fig.\,\ref{Fig:Disp}).
The results of the fits are presented in Table\,\ref{table:Params_fFMR_F}.
\textcolor{newtext}{The final equation for $\nu(L)$ with $f(k)$ in the form (\ref{Eq:MSW_quad_disp}) for Gaussian pulse is:}
\begin{equation}\label{Eq:ChirpParameterFIN}
    \nu(L) = {\frac{\textcolor{newtext}{0.5}\pi F^2 d L}{\textcolor{newtext}{L^2 + d^2(\pi F T_0)^4}}},
\end{equation}
 \textcolor{newtext2}{where $T_0$ is half-duration at $1/e$-intensity point of the wave packet at $L$ = 0 (see Suppl. for details).} 
 
\begin{figure}
\includegraphics[]{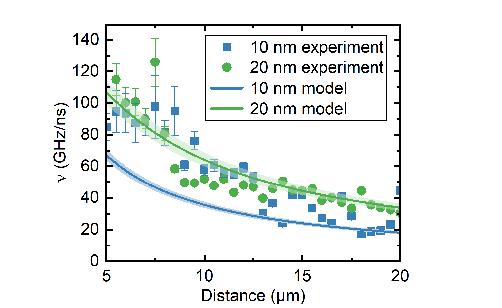}
 \caption{\label{Fig:Chirp} Chirp parameter $\nu$ versus distance for 10 and 20 nm films in HA-configuration.
 Symbols show experimental data, lines -- analytical calculations with Eq.\,(\ref{Eq:ChirpParameterFIN}) with \textcolor{newtext}{$T_0 = 25$ ps, shadowed areas show inaccuracy due to experimental errors of $F$ (Table \ref{table:Params_fFMR_F})}.
 }
 \end{figure}

\textcolor{newtext}{The calculated dependencies $\nu(L)$ with experimental values of $F$ and $T_0 = 25$\,ps are in good agreement with the ones those obtained from} the experimental spectrograms (Fig.\, \ref{Fig:Chirp}).
\textcolor{newtext}{However, the experimentally observed $T_0$ for the MSSW packets is about 200\,ps.
Such deviation could be due to several approximations used in the theoretical approach.}
One of them is $\Delta f \ll f_c$, where $\Delta f$ is the spectral width of the wave packet.
The condition is not fully satisfied in the experiments as we detect MSSW packets with $\Delta f \approx 4$\,GHz at $f_c \approx 9$\,GHz \textcolor{newtext}{in the HA-configuration} (Fig.\,\ref{fig:10nm_experiment}(b-e) and Fig.\,\ref{Fig:Wavelet}(c,d)).
\textcolor{newtext}{The} second approximation is that Eq.\,(\ref{Eq:ChirpParameterFIN}) is obtained for a Gaussian envelope of the packet at the entrance of the media\textcolor{newtext}{, but} the experimental envelope of MSSW is slightly asymmetric in the time domain (Fig.\,\ref{Fig:Wavelet}(a)).
Thus, more comprehensive theoretical analysis of the evolution of SW packets with broadband spectra is needed, but it is out the scope of this letter. 
\textcolor{newtext}{Finally, the extracted dispersion of MSSWs has a rather small term $\frac{\partial^2 f}{\partial k^2}$ (see Fig. \ref{Fig:Disp}), leading to moderate values of chirp detectable reliably at $L$ larger than $10\,\mu$m only.
To make the effect more prominent at smaller distances, it is possible to use magnonic crystals, multilayered structures, or their hybrids \cite{Zakeri_magnonic_crustals_JPCondMat2020,  Sadovnikov_3DCrystal_JMMM2022, Gubbiotti_Magninic3D_APL2021} with a high dispersion of the group velocity of SWs.
On the flipside, the geometrical design of the waveguide paves the way to suppress the dispersion altogether \cite{Divinskiy_DispersionlessPropagation_PhysRevApplied2021, Lake_geometricalControlofSW_APL2021}. }
 
In conclusion, we show experimentally the propagation of optically excited MSSWs in thin iron films with propagation distances of order tens of microns, which may find \textcolor{newtext}{practical} application in magnonic \textcolor{newtext}{circuits} and logic gates.
Additional advantages of iron as a metal for magnonic applications include the relatively easy manufacturing procedure of the thin films and the strong in-plane magneto-crystalline anisotropy.
We use the latter here to tune the character of the MSSW wave packet spectrum evolution in the presence of laser-induced thermal gradients and its chirping with distance.
\textcolor{newtext}{Notably, the chirp of the MSSWs in iron films analogous to the one reported here has been detected in previous experiments with microstripe antennae (Fig. 2 in \cite{Sekiguchi_NPG_Asia_mat_2017}), but was not extensively discussed.}
\textcolor{newtext}{Thus, the presented} analysis of ultrabroadband SW packets \textcolor{newtext}{and their chirping is of current interest for fundamental studies and the design of} future hybrid magnonic-photonic\textcolor{newtext}{-electronic} devices.
\textcolor{newtext}{With this evidence, we ratify} the optical pump-probe technique \textcolor{newtext}{as} a powerful toolbox for the detailed investigation of SWs packets, including: \textcolor{newtext}{capture of the wave} spectral evolution in space, \textcolor{newtext}{its} dispersion law and chirping in the time domain.
The first two options are available with the Brillouin light scattering (BLS) techniques as well \cite{Demokritov_BLS_PhysReports2001,Kalyabin_SSWinTaped_JAP2019}, while the chirping investigation is accessible in BLS with the introduction of phase resolution \textcolor{newtext}{and the} corresponding complication of the measurement scheme \cite{Fohr_phaseBLS_RevSciInst2009, Collet_UltraThinYIG_APL2017}, or, alternatively, with the scanning antennae technique \cite{Annenkov_superdirectional_SW_EPL_2018, Annenkov_analysis_JCTE_2012}.
The ultrafast laser pulses provide all the above possibilities out-of-box with excellent spatial\textcolor{newtext}{-}temporal resolution.

\textcolor{newtext}{
\section*{Supplementary Material}
See supplementary material for the detailed derivation of the equation for chirping parameter $\nu(L)$ and used approximations.
}

\begin{acknowledgments}
The authors thank D.O. Ignatyeva for the fruitful discussion and F. Godejohann for useful tips about the experiment automatization.
A.M.K. acknowledges RSF (grant 20-12-00309),
N.E.Kh. thanks "BASIS" Foundation (grant 19-1-3-42-1) \textcolor{newtext}{for personal support of the theoretical analysis}.
\end{acknowledgments}

\section*{Author declarations}
\subsection*{Conflict of Interest}
The authors have no conflicts to disclose.
\section*{Data availability}
The data that support the findings of this study are available from the corresponding author upon reasonable request.

\bibliography{references}

\end{document}